\UseRawInputEncoding
\pdfoutput=1%For ARXIV
\documentclass[runningheads]{llncs}
\usepackage{graphicx}

\usepackage{tikz}
\usepackage{comment}
\usepackage{amsmath,amssymb} % define this before the line numbering.
\usepackage{color}
\usepackage[accsupp]{axessibility}  % Improves PDF readability for those with disabilities.

\usepackage{adjustbox} 

\usepackage{diagbox}
\usepackage{multirow}
\usepackage[normalem]{ulem}
\usepackage{color}
\usepackage{subfigure}

\usepackage{threeparttable}
\usepackage{makecell}
\usepackage{enumitem}
\usepackage{paralist}
\let\itemize\compactitem
\let\enditemize\endcompactitem
\let\enumerate\compactenum
\let\endenumerate\endcompactenum

\newcommand{\etal}{\textit{et al}.}

\newcommand{\etc}{\textit{etc}. }
\usepackage{gensymb}

\usepackage[pagebackref,breaklinks,colorlinks]{hyperref}
\begin{document}
\pagestyle{headings}
\mainmatter

\makeatletter
\renewcommand*{\@fnsymbol}[1]{\ensuremath{\ifcase#1\or *\or \dagger\or \ddagger\or
   \mathsection\or \mathparagraph\or \|\or **\or \dagger\dagger
   \or \ddagger\ddagger \else\@ctrerr\fi}}
\makeatother

\title{Learning Series-Parallel Lookup Tables for Efficient Image Super-Resolution} 

\newcommand*\samethanks[1][\value{footnote}]{\footnotemark[#1]}

\titlerunning{Learning Series-Parallel Lookup Tables for Efficient Image Super-Resolution}

\author{Cheng Ma\inst{1,2,}\thanks{Equal contribution.} \and
Jingyi Zhang\inst{1,2,}\samethanks \and
Jie Zhou\inst{1,2} \and
Jiwen Lu\inst{1,2,}\thanks{Corresponding author.}
}

\authorrunning{C. Ma*, J.Zhang*, et al.}
\institute{Beijing National Research Center for Information Science and Technology, China \and
Department of Automation, Tsinghua University, China \\
\email{macheng17@tsinghua.org.cn; zhangjy20@mails.tsinghua.edu.cn}\\
\email{\{jzhou, lujiwen\}@tsinghua.edu.cn}
}

%******************
\maketitle

\begin{abstract}
Lookup table (LUT) has shown its efficacy in low-level vision tasks due to the valuable characteristics of low computational cost and hardware independence. 
However, recent attempts to address the problem of single image super-resolution (SISR) with lookup tables are highly constrained by the small receptive field size. Besides, their frameworks of single-layer lookup tables limit the extension and generalization capacities of the model. 
In this paper, we propose a framework of series-parallel lookup tables (SPLUT) to alleviate the above issues and achieve efficient image super-resolution. 
On the one hand, we cascade multiple lookup tables to enlarge the receptive field of each extracted feature vector. 
On the other hand, we propose a parallel network which includes two branches of cascaded lookup tables which process different components of the input low-resolution images. By doing so, the two branches collaborate with each other and compensate for the precision loss of discretizing input pixels when establishing lookup tables. 
Compared to previous lookup table-based methods, our framework has stronger representation abilities with more flexible architectures. 
Furthermore, we no longer need interpolation methods which introduce redundant computations so that our method can achieve faster inference speed. 
Extensive experimental results on five popular benchmark datasets show that our method obtains superior SISR performance in a more efficient way. The code is available at \url{https://github.com/zhjy2016/SPLUT}.
\keywords{Image super-resolution, look-up table, series-parallel network.}
\end{abstract}

\section{Introduction}
\label{sec:intro}

As a fundamental task of computer vision, single image super-resolution (SISR) has attracted lots of research interests and plays an important role in wide applications, such as video surveillance, satellite imaging and high-definition televisions. SISR targets at recovering low-resolution (LR) images to high-resolution (HR) ones by inferring high-frequency details. Along with the rapid development of deep learning techniques, various elaborately designed frameworks~\cite{li2019feedback,lim2017enhanced,zhang2018residual} based on convolutional neural networks (CNNs) have achieved encouraging \linebreak progress in SISR. Most of these frameworks contain a large number of parameters and are time-consuming during testing. While several methods~\cite{lee2020learning,luo2020latticenet} have been proposed to reduce computation costs, they still rely on specific high-performance computing units, for example, GPUs and CPUs. Developing practical and real-time algorithms have been a growing trend in the SISR field.  

Approaches~\cite{zeng2020learning,wang2021learning,wang2021real,jo2021practical} based on lookup tables (LUTs) have emerged in low-level vision tasks, including image enhancement and image super-resolution. These methods employ LUTs to establish the mapping relation between input pixels and the desired output pixels. In the testing phase, only a small number of parameters need storing and the inference processes are liberated from heavy computational burdens by replacing time-consuming calculations with fast memory accesses. As a result, the practicality of this kind of algorithm is significantly improved on mobile devices. 

\begin{figure}[t]
\centering
\subfigure[SR-LUT~\cite{jo2021practical}]{
\centering
\begin{minipage}[b]{0.47\linewidth}
\includegraphics[width=\linewidth]{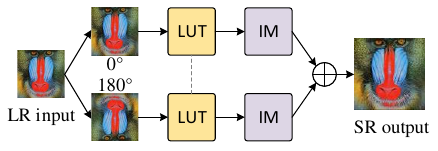}
\end{minipage}
}
\subfigure[SPLUT (Ours)]{
\begin{minipage}[b]{0.47\linewidth}
\centering
\includegraphics[width=\linewidth]{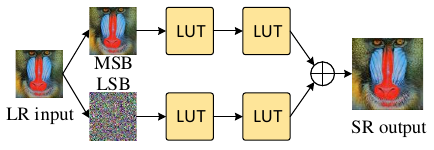}
\end{minipage}
}
\caption{
Comparison of SR-LUT and our SPLUT method. The former utilizes rotational ensemble to enlarge receptive fields from $2\times 2$ to $3\times 3$ and implement interpolation methods (IM) to improve recovery accuracy. The dashed line means weight sharing. For the sake of simplicity, we omit the rotations of $90\degree$ and $270\degree$. In contrast, we stack multiple LUTs to significantly improve the receptive field size and design a new parallel architecture to compensate for the precision loss of discretizing input pixels. 
}
\label{fig:head}
\end{figure}

However, most existing LUT-based approaches only have a single layer of LUTs, which brings some major constraints. 
If $n$-dimensional LUTs ($n$D LUTs) are utilized and the $n$ input entities for query all have $v$ possible values, then the LUT scale is of $v^n$, where $v$ and $n$ are the two pivotal factors. 
While increasing the value of $v$ and $n$ may improve the restoration accuracy, a moderate improvement may lead to a rapid increase of the LUT scale. Thus, $v$ and $n$ are usually set to small values to avoid unbearably large LUTs, which severely limits the further enhancement of recovery abilities. 
In fact, receptive field (RF) size is a vital factor for deep learning and super-resolution. SISR is a well-known ill-posed problem since the same LR input may correspond to various high-resolution outputs with subtle differences.
When we infer the missing details, a large context area on the input image should be considered to accurately capture the semantics and structures. In this way, we can effectively reduce the ambiguity of the estimated results. 
Therefore, how to enlarge receptive fields without exponentially increasing the storage and computation costs of LUTs is still an open issue. 
Besides, due to the design of single-layer LUTs and the limited LUT scale, the extensibility and recovery capacity of existing LUT-based methods are highly constrained. 
Thus, a more powerful and flexible scheme is desired in order to further improve the inference ability of LUT-based methods. 

To mitigate the above issues, we propose to learn series-parallel lookup tables (SPLUT) for SISR, as shown in Fig.~\ref{fig:head}.
We cascade multiple lookup tables so that the query of latter LUTs are based on the outputs of former LUTs. In this way, the receptive field of each feature vector is gradually increased and the final SR results can be determined by larger local patches with more clear context information. For establishing LUTs, existing methods usually discretize the input values for reducing $v$ and apply interpolation algorithms to improve the inference accuracy. However, such operations can only be implemented for small receptive fields. In our framework with large receptive fields, interpolations are inapplicable due to the exponentially increasing computational costs. In order to compensate for the precision loss of the discretized input pixel values, we propose a new parallel network which contains two branches. The first one processes the 4 most significant bits (MSBs) of the original 8-bit pixel values while the second one processes the 4 least significant bits (LSBs). 
The two branches of cascaded LUTs form the framework of series-parallel lookup tables.

In each branch, we introduce three kinds of 4D LUTs whose input values for the query are from different dimensions. We further propose horizontal and vertical aggregation modules to enlarge the RF size of different dimensions. 
In the training procedure, we build a mapping module for each LUT and quantize the intermediate activation so that the mapping relationships of the inputs and outputs can be transferred to the corresponding LUTs.
Different from previous methods~\cite{jo2021practical}, the whole inference procedure of our method only contains retrieval and addition operations without complex multiplications. 
Experimental results on benchmark datasets show that SPLUT can achieve better SR performance on the smartphone platform, which demonstrates the effectiveness and efficiency of our proposed method. 

In summary, the contributions of this work are threefold: 
\begin{enumerate}
\item To the best of our knowledge, we are the first to present cascaded LUTs for enlarging receptive fields in the SISR field.
\item We propose a new parallel network to compensate for the precision loss caused by the discretization when establishing LUTs with large RF sizes.
\item Quantitative and qualitative results show that our method can recover the missing details more precisely and efficiently. The comparison of different SPLUT models verifies the superior extensibility of our proposed method. 
\end{enumerate}
\section{Related Work}

\textbf{Single Image Super-Resolution.}
Non-deep learning methods~\cite{chang2004super,zeyde2010single,timofte2013anchored,timofte2014a+} and deep learning methods~\cite{zhang2018residual,ahn2018fast,park2018srfeat,li2019feedback} have significantly promoted the development of SISR. While recent deep learning methods perform more encouragingly than non-deep learning methods, many of them have a deep neural architecture with redundant parameters, which brings heavy computing costs and makes the training and inference rely on special computing devices.
Therefore, efficient super-resolution has been a prevalent research interest of the community. In this field, methods based on various techniques have been proposed to improve the efficiency of SR algorithms. 
Lee~\etal~\cite{lee2020learning} and Zhang~\etal~\cite{zhang2021data} take advantage of the idea of knowledge distillation to compress the original deep teacher models to small student models with strong representation abilities. 
Wang~\etal~\cite{wang2021exploring} explore the sparsity in image super-resolution by learning sparse masks to identify important regions and unimportant regions in images. 
Mei~\etal~\cite{mei2021image} propose a non-local sparse attention module to achieve efficient and robust long-range modeling. 
Xin~\etal~\cite{xin2020binarized} develop a binary neural network for SR by proposing a bit-accumulation mechanism to improve the precision of the quantized model. 
Some other methods~\cite{luo2020latticenet,song2021addersr,liu2020residual} accomplish efficient SR inference by designing compact neural architectures.
Lee~\etal~\cite{lee2020journey} search for appropriate architectures for both the generator and the discriminator by a neural architecture search approach. 
However, most of these methods are still based on convolutional layers and thus lack practicability on mobile devices.

\textbf{Lookup Table.} Lookup tables (LUTs) replace complex computations by simple and fast retrieval operations so that the efficiency of algorithms can be significantly improved.  LUTs are widely used in a number of applications, such as numerical computation~\cite{845004,465093}, video coding~\cite{5473028,6253235}, pedestrian detection~\cite{7492605}, RGB-to-RGBW conversion~\cite{7236906}, \etc
Besides, LUT is a classic and prevalent pixel adjustment tool in camera imaging pipeline~\cite{zeng2020learning} and photo editing software since it can easily manipulate the appearance of an image, such as color, exposure, saturation, \etc Recently deep learning methods based on LUTs have also emerged in low-level vision tasks~\cite{zeng2020learning,wang2021learning,wang2021real,jo2021practical}.
In the image enhancement field, Zeng~\etal~\cite{zeng2020learning} first propose image-adaptive 3D lookup tables and achieve high-performance photo enhancement. On this basis, Wang~\etal~\cite{wang2021real} consider spatial information and further propose learnable spatial-aware 3D lookup tables. Wang~\etal~\cite{wang2021learning} model local context cues and propose pixel-adaptive lookup table weights for portrait photo retouching. 
As for the super-resolution field, Jo~\etal~\cite{jo2021practical} has developed SR-LUT by establishing the correspondence of LR input patterns and HR output patterns. 
However, as mentioned above, the RF size and the extensibility of LUT-based methods are still limited. 

\section{Method}
\subsection{Network Architecture}
\begin{figure*}[t]
\centering
\subfigure[Overall framework]{
\begin{minipage}[b]{\linewidth}
\centering
\includegraphics[width=0.95\linewidth]{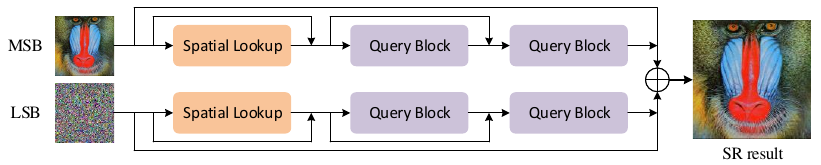}
\end{minipage}
}

\subfigure[Query block]{
\begin{minipage}[b]{0.78\linewidth}
\centering
\includegraphics[width=\linewidth]{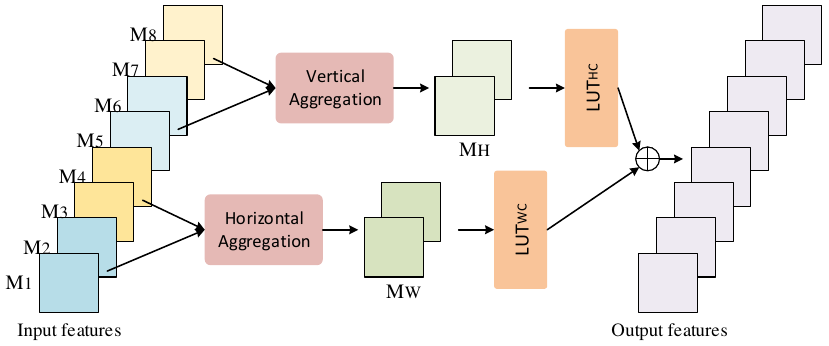}
\end{minipage}
}
\subfigure[LUTs]{
\begin{minipage}[b]{0.15\linewidth}
\centering
\includegraphics[width=\linewidth]{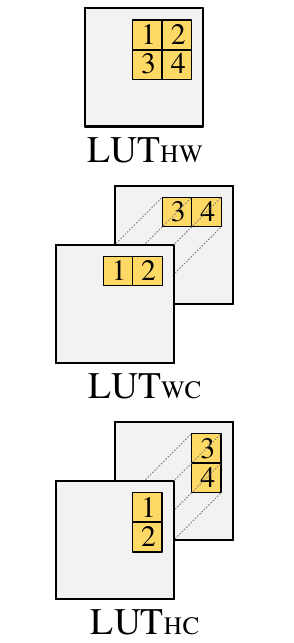}
\end{minipage}
}
\caption{
Details of the proposed SPLUT method. (a) The overall framework of our method. The input LR images are split into $I_{MSB}$ and $I_{LSB}$, which are fed into two parallel branches, respectively. Each branch includes cascaded LUTs to extend receptive fields. (b) We take SPLUT-M with $C_f=8$ as the example and display the details of the proposed query block, which further enlarge the RF size by aggregation modules and different kinds of LUTs. $\rm LUT_{WC}$ and $\rm LUT_{HC}$ can also model the correlations between different channels. (c) Illustration of different LUTs: $\rm LUT_{HW}$, $\rm LUT_{WC}$ and $\rm LUT_{HC}$. Their input values for the query are in different dimensions. 
}
\label{fig:frame}
\end{figure*}

Given an input LR image $I^{LR}$, our goal is to recover the missing details and yield the SR image $I^{SR}$ which is as similar as possible to the HR image $I^{HR}$. As shown in Fig.~\ref{fig:frame} (a), we design a series-parallel lookup table (SPLUT) network which contains two parallel branches processing different components of $I^{LR}$. We treat the RGB channels equally and separate the original input pixels with 8-bit values into two maps, $I_{MSB}$ with 4 most significant bits (MSBs) and $I_{LSB}$ with 4 least significant bits (LSBs). The two parallel branches take $I_{MSB}$ and $I_{LSB}$ as inputs, respectively. Then we merge the outputs of the two parallel branches to compensate for the loss of quantization when establishing LUTs. In this way, the super-resolution capacities can be significantly enhanced. In each branch, there is a spatial lookup block, query blocks and skip connections. The spatial lookup block and the query blocks increase the RF size of extracted features gradually. The query blocks include horizontal and vertical aggregation modules which enlarge the RF size by the width and height dimensions, respectively. During training, we replace each LUT with a mapping block which is built on convolutional layers. Then we establish LUTs according to the mapping relations of the inputs and outputs of these mapping blocks. During inference, we retrieve the outputs of each LUT according to the indices computed by the input patterns, which are defined as the combinations of the $n$ input entities for the query. 

The LUT scale is mainly influenced by three factors, the number of pixels for retrieval $n$, the number of possible pixel values $v$, and the length of output vectors $c$. Then the LUT size can be computed by $v^n\cdot c$. In previous work, $n$ is usually not more than 4 to keep a small LUT scale. The original pixels with 256 different values are also discretized to obtain $v=16$ or $v=32$ bins for retrieval. The choice of $c$ is determined by the practical tasks. For generating the output of $\times 4$ super-resolution, $c$ is set to 16. The increase of $n$ and $v$ results in a significant increase in LUT scales while $c$ only brings a linear growth of LUT scales.  In our framework, we set $n=4$ and $v=16$ for all the LUTs so that $v^n=65536$ is a relatively small constant. Different from previous methods which only contain a single layer of LUTs with a small RF size and lack the model extensibility, we cascade multiple LUTs to improve the RF size and flexibly control the trade-off between efficiency and accuracy by changing the network depth and the channel number of intermediate features $C_f$. In practice, we design three models, SPLUT-S, SPLUT-M and SPLUT-L with $C_f=4$, $C_f=8$ and $C_f=16$, respectively. 

Since $n$ is set to 4, the indices for retrieving LUTs are computed by 4 adjacent entities. We design 3 kinds of LUTs whose input patterns are of different dimensions. For an intermediate feature, there are mainly three dimensions, W, H, and C, representing width, height, and channel, respectively. The 3 kinds of designed LUTs are $\rm LUT_{WH}$, $\rm LUT_{HC}$ and $\rm LUT_{WC}$, as depicted in Fig.~\ref{fig:frame} (c). The input pattern of $\rm LUT_{WH}$ is a $2\times 2$ area in the spatial dimensions. The $2\times 2$ input pattern of $\rm LUT_{HC}$ is along the height and channel dimensions while that of $\rm LUT_{WC}$ is along the width and channel dimensions. These LUTs can capture local dependency and enlarge receptive fields of different dimensions. In our framework, they are placed in different modules for specific functions. 
Next, we take the model SPLUT-M as an example to describe the details of each component of our framework.

\textbf{Spatial LUT Block.} The two branches of $I_{MSB}$ and $I_{LSB}$ have similar architectures. In the beginning, spatial correlations are more important than channel correlations. Hence, we use a spatial LUT block to exploit the spatial dependency of neighboring pixels in the input images. In this block, we employ reflection padding to keep the spatial dimensions unchanged after retrievals.  

\begin{figure}[t]
\centering
\includegraphics[width=0.9\linewidth]{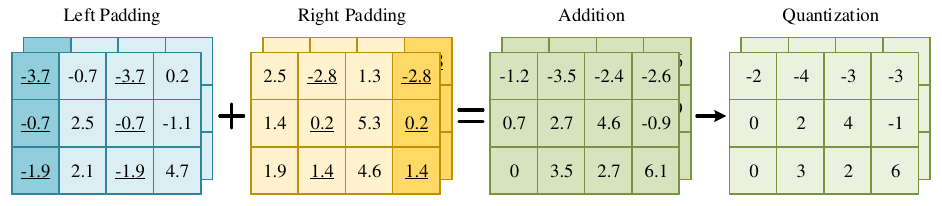}
\caption{
Illustration of the proposed horizontal aggregation module. The underlined numbers in dark squares represent the results of the reflection padding operations. After adding the two padded feature maps together, the receptive field of the obtained features is enlarged on the width dimension. 
}
\label{fig:aggregation}
\end{figure}

\textbf{Query Blocks.} Following the spatial LUT block, there are two query blocks. The details of the query block are shown in Fig.~\ref{fig:frame} (b). For the model of SPLUT-M, $C_f$ is set to 8 and thus we have 8 intermediate feature maps, named $M_1, ... , M_8$. We split them into 4 groups, each adjacent two in one group. 
The first two groups are fed into the horizontal aggregation module while the last two are fed into the vertical aggregation module.
The two modules enlarge the RF size by the width dimension and the height dimension, respectively. $M_W$ and $M_H$ are obtained by the two modules and they both have 2 channels. Since only exploring spatial information severely affects the representation ability of the network, we use $\rm LUT_{WC}$ and $\rm LUT_{HC}$ to model the correlations between different channels. We retrieve the output of a width-channel $\rm LUT_{WC}$ by computing the query indices according to the input patterns on $M_W$. Similarly, we retrieve $\rm LUT_{HC}$ according to $M_H$. Finally, the outputs of two kinds of LUTs are added to get the output of the query block. 

\textbf{Aggregation Modules.} Here we describe the details of the horizontal and vertical aggregation modules. As shown in Fig.~\ref{fig:aggregation},  we take the horizontal aggregation module as an example. The two input feature maps both have two channels and have the same receptive fields. First, we pad one feature map on the left by reflection padding and pad the other one on the right. Then we obtain two feature maps whose receptive fields have a shift of one pixel along the width dimension. After merging the two feature maps by addition, the obtained feature map has a larger spatial receptive field. In order to transfer the real-value responses to the query indices for the following LUTs, we need to quantize the real values to form $v=16$ discrete values. Specifically, we set the quantization interval to 1 so that the quantization can be achieved by a simple rounding operation. By doing this, we avoid complex multiplication computations and improve the efficiency of the proposed module. The operations are similar for the vertical aggregation module. Differently, vertical aggregation enlarges the receptive field along the height dimension. 

\textbf{Parallel Branches.} In prior arts~\cite{zeng2020learning,jo2021practical,wang2021real,wang2021learning}, pixel values are quantized for reducing the possible values and decreasing LUT scales. However, the original continuously changing pixels become discrete, which may cause blocking effects in the SR results. Therefore, interpolation algorithms~\cite{kasson1995performing} are usually applied to smooth the output textures. 
However, they introduce additional multiplications and comparison operations. 
Moreover, these algorithms are only available for LUTs with a small RF size. If we use $r$ to represent the RF size of a feature, then $2^r$ nearest bounding vertices need considering for interpolating the retrieval results. In our framework with a large RF size, such a computation complexity is unacceptable. We propose a new parallel framework to alleviate this issue. The framework includes two branches with the same architecture of cascaded LUTs. One branch processes $I_{MSB}$ and mainly focuses on capturing context semantic information. The other branch processes $I_{LSB}$ and provides high-frequency details. By Merging the outputs of the two branches, we are able to compensate for the loss of quantization when establishing LUTs and hence boost SR performance.

\begin{figure*}[t]
\centering
\includegraphics[width=0.95\linewidth]{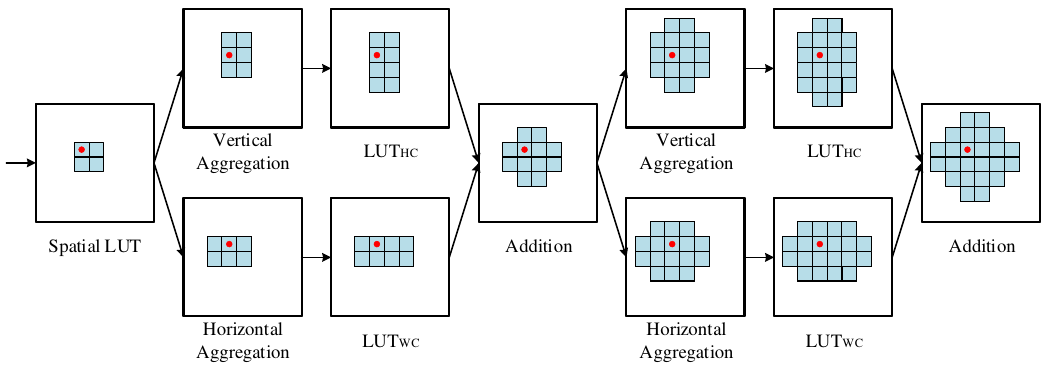}
\caption{
The visualization of receptive fields with respect to the feature marked in red after each module or operation. Horizontal aggregation modules and $\rm LUT_{WC}$ increase the RF size by the width dimension while vertical aggregation modules and $\rm LUT_{HC}$ increase the RF size by the height dimension. The addition operations further enlarge receptive fields by fusing the two different areas of receptive fields.
}
\label{fig:receptive}
\end{figure*}

\textbf{Skip Connections.} In order to improve the representation abilities of the network, we store low-precision real numbers in LUTs. Since the above mentioned operations of quantization and index computation sacrifice the precision of intermediate features, we introduce skip connections to fuse the real-value inputs and the retrieval outputs to improve the precision. Besides, identity mapping~\cite{he2016deep} is a pivotal component of  SR networks. Thus we adopt a skip connection between the input image and the output of the last query block to simplify optimization and enhance recovery accuracy.

\subsection{Training Strategy}

For training the network, we replace the LUTs by mapping modules, which are comprised of a convolutional layer with a kernel size of $k_h\times k_w$, GELU~\cite{hendrycks2016gaussian} layers, and $1\times 1$ convolutional layers. 
% The architecture is displayed in Fig.~\ref{fig:train}. 
All mapping modules output feature maps with $C_f$ channels except the last one. The last mapping module outputs $C_{sr}=s^2$ channels where $s$ is the upscaling factor. A pixel-shuffle layer~\cite{shi2016real} maps the outputs with 16 channels to the final results for $\times 4$ SR. In mapping modules of different LUTs, $k_h$ and $k_w$ are different. For the spatial LUT block, the input channel number is 1 and $k_h=k_w=2$. For $\rm LUT_{HC}$ and $\rm LUT_{WC}$, the input channel number is 2. $k_h=2$ and $k_w=1$ for $\rm LUT_{HC}$ while $k_h=1$ and $k_w=2$ for $\rm LUT_{WC}$. The consecutive $1\times 1$ convolutions followed by GELU layers strengthen the nonlinearity and representative abilities of the mapping modules. 
We jointly train the MSB and LSB branches by imposing Mean Squared Error (MSE) loss on the final SR outputs. 
For the quantized activations, we use the identity straight-through estimator (STE)~\cite{yin2019understanding} to achieve end-to-end back-propagation. 

\subsection{Analysis of SPLUT}

\textbf{Receptive Field Size.} We visualize the changes of RF sizes through the whole framework in Fig.~\ref{fig:receptive}. After the first spatial LUT block, each feature has an RF size of $2\times 2$. Then in the first query block, the horizontal aggregation module and $\rm LUT_{WC}$ enlarge the RF size along the width dimension. The vertical aggregation module and $\rm LUT_{HC}$ enlarge the RF size along the height dimension. After fusing the two outputs by addition, we get an RF size of 12. In the second query block, we implement similar operations and further increase the RF size. Finally, we get an RF size of 24, which is much bigger than the RF size of $3\times 3=9$ in SR-LUT~\cite{jo2021practical} by rotational ensemble trick. 

\textbf{Computational Cost.} In SR-LUT~\cite{jo2021practical}, rotational ensemble is proposed to extend the RF size of $2\times 2$ to $3\times 3$. The computational burdens of SR-LUT mainly include image rotation, retrieval of 4 input images with different orientations, and interpolation methods which contain heavy multiplication and comparison operations. While our SPLUT model has more LUTs, we do not need the rotation and interpolation operations. 
Therefore, our method can achieve a faster inference speed than SR-LUT. 
\section{Experiments}

\subsection{Implementation Details}
\textbf{Datasets and Metrics}. 
We train the proposed serial-parallel lookup table (SPLUT) model on the DIV2K dataset~\cite{agustsson2017ntire} and evaluate the effectiveness of our method on 5 widely used benchmarks: Set5~\cite{bevilacqua2012low}, Set14~\cite{zeyde2010single},BSD100~\cite{BSD100}, Urban100~\cite{huang2015single}  and Manga109~\cite{mtap_matsui_2017}.
We focus on the upscaling factor of $\times  4$ in our experiments. 
We use Peak Signal-to-Noise Ratio (PSNR) and structural similarity index (SSIM)~\cite{wang2004image} as the evaluation metrics for prediction accuracy. 
To compare the computation efficiency, we measure and report the runtime of super-resolving $320\times180$ LR images on mobile phones. 

\textbf{Training Setting}. We design three SPLUT models with different model sizes, namely SPLUT-S, SPLUT-M, and SPLUT-L. 
The three models have the same architecture depicted in Fig.~\ref{fig:framework} (a). 
The difference between the three models lies in the value of $C_f$, the number of lookup tables per query block and the grouping strategy of input feature maps. 
We have introduced the details of SPLUT-M. The details of the other two models are described in the supplementary material.  
We train SPLUT models with PyTorch~\cite{PyTorch} on Nvidia 2080Ti GPUs. We use Adam Optimizer~\cite{Adam} with $\beta_1=0.9, \beta_2=0.999$ and $\epsilon=1\times10^{-8}$ to jointly train the MSB and LSB branches. The learning rate is set to $10^{-3}$. 
We randomly crop LR images into $48\times48$ patches with a mini-batch size of 32. We enhance the dataset by randomly rotating and flipping.

\subsection{Results and Analyses}

%Comparison
\begin{table}[t]
\scriptsize
\begin{center}
\caption{Quantitative comparisons of different SR methods on 5 benchmark datasets. The best results among LUT-based methods are \textbf{highlighted}. Running time is measured by super-resolving 320 × 180 LR images on the mobile phone. * represents the running time is measured on computer CPUs. Size denotes the storage space or the  parameter number of each model. }
\label{tab:SOTA}
\renewcommand\arraystretch{1.5} % set height
\setlength{\tabcolsep}{1pt}
{ % set width
    \begin{tabular}{c|c|c|cc|cc|cc|cc|cc}
    \hline
    \cline{1-13} 
    \multirow{2}{*}{Method} & \multirow{2}{*}{Time} & \multirow{2}{*}{Size} & \multicolumn{2}{c|}{Set5}                                  & \multicolumn{2}{c|}{Set14}                                 & \multicolumn{2}{c|}{BSDS100}                               & \multicolumn{2}{c|}{Urban100}                              & \multicolumn{2}{c}{Manga109}             \\ \cline{4-13} 
                            &                      &                & \multicolumn{1}{c|}{PSNR}              & SSIM              & \multicolumn{1}{c|}{PSNR}              & SSIM              & \multicolumn{1}{c|}{PSNR}              & SSIM              & \multicolumn{1}{c|}{PSNR}              & SSIM              & \multicolumn{1}{c|}{PSNR}              & SSIM     \\ 
                            \hline
                            % \hline
    NE+LLE                & 7016ms*                  & 1.434MB                & \multicolumn{1}{c|}{29.62}             & 0.840            & \multicolumn{1}{c|}{26.82}             & 0.735            & \multicolumn{1}{c|}{26.49}             & 0.697            & \multicolumn{1}{c|}{23.84}             & 0.694            & \multicolumn{1}{c|}{26.10}             & 0.820   \\
    Zeyde~\etal                & 8797ms*                  & 1.434MB                & \multicolumn{1}{c|}{26.69}             & 0.843            & \multicolumn{1}{c|}{26.90}             & 0.735            & \multicolumn{1}{c|}{26.53}             & 0.697            & \multicolumn{1}{c|}{23.90}             & 0.696            & \multicolumn{1}{c|}{26.24}             & 0.824   \\
    ANR                     & 1715ms*                  & 1.434MB                & \multicolumn{1}{c|}{29.70}             & 0.842            & \multicolumn{1}{c|}{26.86}             & 0.737            & \multicolumn{1}{c|}{26.52}             & 0.699            & \multicolumn{1}{c|}{23.89}             & 0.696            & \multicolumn{1}{c|}{26.18}             & 0.821   \\
    A+                      & 1748ms*                  & 15.17MB               & \multicolumn{1}{c|}{30.27}             & 0.860            & \multicolumn{1}{c|}{27.30}             & 0.750            & \multicolumn{1}{c|}{26.73}             & 0.709            & \multicolumn{1}{c|}{24.33}             & 0.719            & \multicolumn{1}{c|}{26.91}             & 0.848   \\
    CARN-M                  & 4955ms                   & 1.593MB                  & \multicolumn{1}{c|}{31.82}             & 0.890            & \multicolumn{1}{c|}{28.29}             & 0.775            & \multicolumn{1}{c|}{27.42}             & 0.730            & \multicolumn{1}{c|}{25.62}             & 0.769            & \multicolumn{1}{c|}{29.85}             & 0.899   \\
    % IMDN                      & 3987ms                  & 2.74 MB                 & \multicolumn{1}{c|}{32.21}              & 0.895            & \multicolumn{1}{c|}{28.58}             & 0.781         & \multicolumn{1}{c|}{27.56}             & 0.735            & \multicolumn{1}{c|}{26.04}             & 0.784            & \multicolumn{1}{c|}{30.45}             & 0.908 \\ 
    
    FMEN                         & 3101ms                   & 1.395MB            & \multicolumn{1}{c|}{32.24}         & 0.896                & \multicolumn{1}{c|}{28.70}         & 0.784                & \multicolumn{1}{l|}{27.63}         &0.738                 & \multicolumn{1}{c|}{26.28}         & 0.791                & \multicolumn{1}{c|}{30.70}         & 0.911\\ 
    
    % EDSR                      & 52917ms                  & 164.4MB                 & \multicolumn{1}{c|}{32.46}             & 0.897            & \multicolumn{1}{c|}{28.80}             & 0.788            & \multicolumn{1}{c|}{27.71}             & 0.742            & \multicolumn{1}{c|}{26.64}             & 0.803            & \multicolumn{1}{c|}{31.02}             & 0.915   \\
    RRDB                   & 31717ms                  & 63.83MB            & \multicolumn{1}{c|}{32.60}             & 0.900            & \multicolumn{1}{c|}{28.88}             & 0.790            & \multicolumn{1}{c|}{27.76}             & 0.743            & \multicolumn{1}{c|}{26.73}             & 0.807            & \multicolumn{1}{c|}{31.16}             & 0.916   \\ \hline

    SR-LUT                   & 279ms                    & \textbf{1.274M}       & \multicolumn{1}{c|}{29.82}             & 0.848            & \multicolumn{1}{c|}{27.01}             & 0.736            & \multicolumn{1}{c|}{26.53}             & 0.695            & \multicolumn{1}{c|}{24.02}             & 0.699            & \multicolumn{1}{c|}{26.80}              & 0.838   \\
    SPLUT-S                  & \textbf{242ms}           & 5.5M                  & \multicolumn{1}{c|}{30.01}             & 0.852            & \multicolumn{1}{c|}{27.20}             & 0.743            & \multicolumn{1}{c|}{26.68}             & 0.702            & \multicolumn{1}{c|}{24.13}             & 0.706            & \multicolumn{1}{c|}{27.00}             & 0.843   \\
    SPLUT-M                  & 265ms                    & 7M                    & \multicolumn{1}{c|}{30.23}             & 0.857            & \multicolumn{1}{c|}{27.32}             & 0.746            & \multicolumn{1}{c|}{26.74}             & 0.704            & \multicolumn{1}{c|}{24.21}             & 0.709            & \multicolumn{1}{c|}{27.20}             & 0.848   \\
    SPLUT-L                  & 545ms                    & 18M                   & \multicolumn{1}{c|}{\textbf{30.52}}    & \textbf{0.863}   & \multicolumn{1}{c|}{\textbf{27.54}}    & \textbf{0.752}   & \multicolumn{1}{c|}{\textbf{26.87}}    & \textbf{0.709}   & \multicolumn{1}{c|}{\textbf{24.46}}    & \textbf{0.719}   & \multicolumn{1}{c|}{\textbf{27.70}}    & \textbf{0.858}   \\ 
    \hline
    \cline{1-13} 
    \end{tabular}}
\end{center}
\end{table}

\textbf{Quantitative Comparison}. We compare our method with SR methods based on sparse coding which include NE+LLE~\cite{chang2004super}, Zeyde~\etal~\cite{zeyde2010single}, ANR~\cite{timofte2013anchored} and A+~\cite{timofte2014a+}, SR methods based on deep learning including CARN-M~\cite{ahn2018fast}, FMEN~\cite{du2022fast}
% , EDSR~\cite{lim2017enhanced}
and RRDB~\cite{wang2018esrgan}, and SR method based on LUTs, SR-LUT~\cite{jo2021practical}. 
Since the source code of SR-LUT~\cite{jo2021practical} is not released, we reproduce the SR-LUT algorithm and compare our method with it under the same environment. 
Since the implementation of sparse coding based methods~\cite{chang2004super,zeyde2010single,timofte2013anchored,timofte2014a+} rely on Matlab, we evaluate these methods on the CPUs of computers which may be faster than mobile phones. 
The quantitative comparisons are shown in Table~\ref{tab:SOTA}. 
As observed, our SPLUT models achieve much faster inference than both sparse coding based methods and deep learning based methods. SPLUT-S, SPLUT-M and SPLUT-L all obtain higher PSNR and SSIM than NE+LLE, Zeyde~\etal and ANR. The SPLUT-M is comparable to A+ and SPLUT-L is superior to A+ on all benchmarks. While deep learning based methods have the best PSNR and SSIM performance, their inference speed is much slower than our method. As a method based on LUTs, SR-LUT is much faster than the other compared methods. However, it is still slower than our SPLUT-S and SPLUT-M methods. Besides, SPLUT models all outperform SR-LUT by a large margin on PSNR and SSIM metrics. Our SPLUT-M model achieves a better trade-off between efficiency and accuracy. Compared to SR-LUT, SPLUT-M improves PSNR by 0.4 dB on Set5 and Manga109 in a faster speed. 
By comparing model sizes, we see that our SPLUT method only brings a linear increase in storage costs. We believe the LUT size of our method is acceptable for current mobile phones. Therefore, we regard the runtime as a more important factor for evaluating efficiency. 
Besides, SPLUT-L presents more powerful SR abilities than SPLUT-S and SPLUT-M. These comparisons verify that our SPLUT framework is more powerful and more flexible than the previous framework of single-layer LUTs whose scale increases exponentially with RF sizes. 

\begin{figure*}[t]
\newlength\fsdurthree
\setlength{\fsdurthree}{0.5mm}
\centering
\begin{adjustbox}{valign=t}
  \scriptsize
  \begin{tabular}{ccccccc}
      &  Bicubic& A+ & SRLUT & SPLUT-S& SPLUT-M & HR \\
    
    \includegraphics[width=0.123\textwidth]{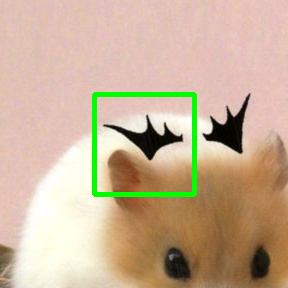} &
    \includegraphics[width=0.123\textwidth]{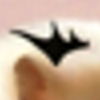} &
    \includegraphics[width=0.123\textwidth]{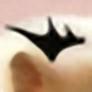} &
    \includegraphics[width=0.123\textwidth]{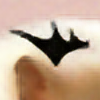} &
    \includegraphics[width=0.123\textwidth]{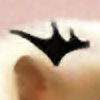} &
    \includegraphics[width=0.123\textwidth]{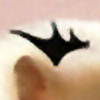} &
    \includegraphics[width=0.123\textwidth]{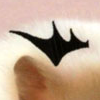} \\

    \includegraphics[width=0.123\textwidth]{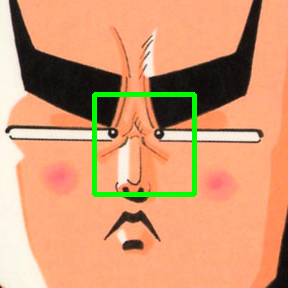} &
    \includegraphics[width=0.123\textwidth]{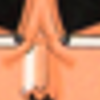} &
    \includegraphics[width=0.123\textwidth]{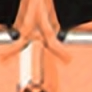} &
    \includegraphics[width=0.123\textwidth]{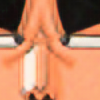} &
    \includegraphics[width=0.123\textwidth]{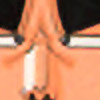} &
    \includegraphics[width=0.123\textwidth]{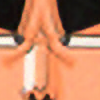} &
    \includegraphics[width=0.123\textwidth]{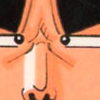} \\

    \includegraphics[width=0.123\textwidth]{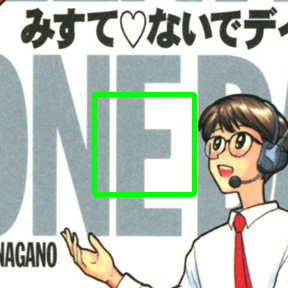} &
    \includegraphics[width=0.123\textwidth]{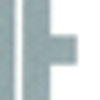} &
    \includegraphics[width=0.123\textwidth]{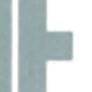} &
    \includegraphics[width=0.123\textwidth]{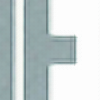} &
    \includegraphics[width=0.123\textwidth]{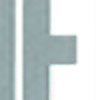} &
    \includegraphics[width=0.123\textwidth]{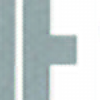} &
    \includegraphics[width=0.123\textwidth]{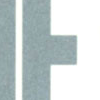} \\

    \includegraphics[width=0.123\textwidth]{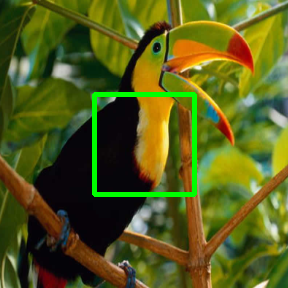} &
    \includegraphics[width=0.123\textwidth]{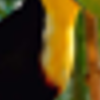} &
    \includegraphics[width=0.123\textwidth]{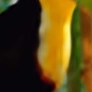} &
    \includegraphics[width=0.123\textwidth]{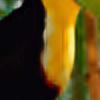} &
    \includegraphics[width=0.123\textwidth]{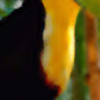} &
    \includegraphics[width=0.123\textwidth]{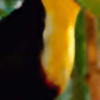} &
    \includegraphics[width=0.123\textwidth]{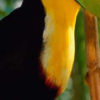} \\
    
    \includegraphics[width=0.123\textwidth]{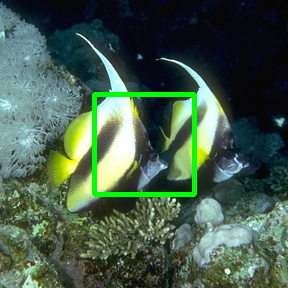} &
    \includegraphics[width=0.123\textwidth]{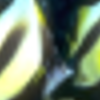} &
    \includegraphics[width=0.123\textwidth]{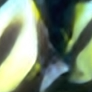} &
    \includegraphics[width=0.123\textwidth]{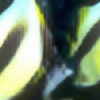} &
    \includegraphics[width=0.123\textwidth]{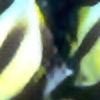} &
    \includegraphics[width=0.123\textwidth]{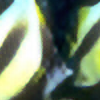} &
    \includegraphics[width=0.123\textwidth]{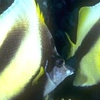} \\

  \end{tabular}
\end{adjustbox}
\caption{
	Qualitative Comparisons of bicubic interpolation, A+~\cite{timofte2014a+}, SRLUT~\cite{jo2021practical}, our SPLUT method and HR images. The results show our method can generate sharp edges without severe artifacts. 
}
\label{fig:all_visual}
\end{figure*}

\textbf{Qualitative Comparison}. Fig.~\ref{fig:all_visual} illustrates the qualitative comparisons of Bicubic interpolation, A+, SR-LUT, our SPLUT models, and ground-truth images. 
We can see SR-LUT fails to present natural details for sharp edges. In some areas with continuously changing colors, there are often blocking artifacts. 
In the third row, the SR results of SR-LUT have severe ringing artifacts near edges. 
While A+ introduces fewer artifacts than SR-LUT, it may generate more blurry edges, as shown in the second row. 
On the contrary, our SPLUT models restore more natural textures. 
It can be seen that the expansion of the receptive field in SPLUT helps the network grasp the texture and structure information of context regions to achieve better reconstruction accuracy.

%Comparison interpret
\begin{table}[t]
\scriptsize
\begin{center}
\caption{Comparison of parallel network and interpolation methods. The results show OBM w/o interpolation and the models with interpolation methods cannot achieve comparable performance to our SPLUT method. }
\label{tab:interpret}
\renewcommand\arraystretch{1.5} % set height
\setlength{\tabcolsep}{3pt}{ % set width
    \begin{tabular}{c|c|cc|cc}
    \hline
    \cline{1-6}
    \multirow{2}{*}{method}    & \multirow{2}{*}{Size} & \multicolumn{2}{c|}{Set5}           & \multicolumn{2}{c}{Set14}           \\ \cline{3-6} 
                            &                       & \multicolumn{1}{c|}{PSNR}  & SSIM   & \multicolumn{1}{c|}{PSNR}  & SSIM   \\ 
                            \hline
                            % \hline
     OBM w/o interpolation            & 3.5M                  & \multicolumn{1}{c|}{27.24} & 0.8217 & \multicolumn{1}{c|}{25.30} & 0.7175 \\ 
    Tail-layer Interpolation        & 3.5M                  & \multicolumn{1}{c|}{27.24} & 0.8217 & \multicolumn{1}{c|}{25.30} & 0.7175  \\ 
    First-layer Interpolation     & 3.5M                  & \multicolumn{1}{c|}{27.24} & 0.8218 & \multicolumn{1}{c|}{25.30} & 0.7176 \\
        SPLUT              & 7M                    & \multicolumn{1}{c|}{\textbf{30.23}} & \textbf{0.8567} & \multicolumn{1}{c|}{\textbf{27.32}} & \textbf{0.7460} \\
    \hline
    \cline{1-6}
    \end{tabular}}
\end{center}
\end{table}

\textbf{Ablation: Parallel Network vs. Interpolation}. We take SPLUT-M as the baseline model and further investigate the effectiveness of our proposed parallel network by comparing it with interpolation algorithms. 
Specifically, we remain only one branch of the SPLUT model and use full-precision LR images as inputs to train this model. In the inference phase, we follow SR-LUT~\cite{jo2021practical} to extract $I_{MSB}$ for retrieval and store $I_{LSB}$ for interpolation. We call the one-branch model without interpolation ``OBM w/o interpolation''. 
Since the cascaded LUTs in this model brings a large RF size of $r$, it is intractable to consider all bounding vertices and simply implement interpolations due to the computational complexity of $2^r$.
To improve the SR accuracy of this model, we design two interpolation methods for the input images. 
For a position of $(x,y)$, SR-LUT implements 4-simplex interpolation for 4D LUTs by exploring the relation of $I_{LSB}^{(x,y)}$, $I_{LSB}^{(x+1,y)}$, $I_{LSB}^{(x,y+1)}$ and $I_{LSB}^{(x+1,y+1)}$. A weighted sum of the retrieval results for the 16 bounding vertices is computed as the final output since SR-LUT has only one layer of LUT. In our methods, we also utilize $I_{LSB}$ for interpolation but we cannot get the final SR output by directly fusing the retrieval results of the first layer of spatial LUT since we still have other following LUTs for retrieval. Therefore, we concatenate the 16 bounding vertices of all input pixels to form 16 index maps, which reduce the complexity of $2^r$ to $2^4$. In our first interpolation method, we take these 16 index maps as the inputs to the spatial lookup blocks and get 16 SR results through the whole network. We interpolate the 16 SR results by $I_{LSB}$ using the 4-simplex method. We call this method tail-layer interpolation. In our second method, we feed the 16 index maps to the spatial lookup blocks and get 16 intermediate feature maps. We fuse the 16 feature maps by $I_{LSB}$ to get one feature map. By feeding the feature map to the following layers, we can get a final SR output. We call this method first-layer interpolation. More implementation details are described in the supplementary. 
The results are shown in Table~\ref{tab:interpret}. Our SPLUT with the parallel network achieves an improvement of more than 2.9 dB  over the two interpolation methods and OBM w/o interpolation. 
This proves that applying interpolation methods fails to compensate for the precision loss, which is caused by discretizing pixel values when establishing lookup tables. In contrast, our parallel network is superior to interpolation algorithms in compensating for the precision loss for large RF sizes. It can also be inferred that our parallel network is inherently robust to different receptive fields.

%Sampling Interval
\begin{table}[t]
\scriptsize
\begin{center}
\caption{Comparison of SPLUT models with different quantization precision $v_f$ of intermediate features. We choose $v_f=16$ as the optimal setting considering accuracy and efficiency. }
\label{tab:Sampling Interval}
\renewcommand\arraystretch{1.5} % set height
\setlength{\tabcolsep}{3pt}{ % set width
\begin{tabular}{c|c|cc|cc|cc|cc|cc}
\hline
\cline{1-12} 
\multirow{2}{*}{$v_f$} & \multirow{2}{*}{Size}      & \multicolumn{2}{c|}{Set5}           & \multicolumn{2}{c|}{Set14}          & \multicolumn{2}{c|}{BSDS100}        & \multicolumn{2}{c|}{Urban100}       & \multicolumn{2}{c}{Manga109}        \\ \cline{3-12} 
                    &          & \multicolumn{1}{c|}{PSNR}  & SSIM   & \multicolumn{1}{c|}{PSNR}  & SSIM   & \multicolumn{1}{c|}{PSNR}  & SSIM   & \multicolumn{1}{c|}{PSNR}  & SSIM   & \multicolumn{1}{c|}{PSNR}  & SSIM   \\ 
                    \hline
8                   & 1.375M  & \multicolumn{1}{c|}{29.95} & 0.849 & \multicolumn{1}{c|}{27.13} & 0.738 & \multicolumn{1}{c|}{26.63} & 0.698 & \multicolumn{1}{c|}{24.05} & 0.702 & \multicolumn{1}{c|}{26.83} & 0.839 \\
12                  & 2.898M  & \multicolumn{1}{c|}{30.11} & 0.854 & \multicolumn{1}{c|}{27.26} & 0.743 & \multicolumn{1}{c|}{26.71} & 0.702 & \multicolumn{1}{c|}{24.15} & 0.706 & \multicolumn{1}{c|}{27.06} & 0.844 \\
16                  & 7M      & \multicolumn{1}{c|}{30.23} & 0.857 & \multicolumn{1}{c|}{27.32} & 0.746 & \multicolumn{1}{c|}{26.74} & 0.704 & \multicolumn{1}{c|}{24.21} & 0.709 & \multicolumn{1}{c|}{27.20} & 0.848 \\
20                  & 15.648M & \multicolumn{1}{c|}{30.22} & 0.856 & \multicolumn{1}{c|}{27.31} & 0.746 & \multicolumn{1}{c|}{26.74} & 0.704 & \multicolumn{1}{c|}{24.22} & 0.710 & \multicolumn{1}{c|}{27.25} & 0.848 \\
24                  & 31.375M & \multicolumn{1}{c|}{30.25} & 0.858 & \multicolumn{1}{c|}{27.34} & 0.747 & \multicolumn{1}{c|}{26.75} & 0.705 & \multicolumn{1}{c|}{24.24} & 0.711 & \multicolumn{1}{c|}{27.24} & 0.849 \\
\hline
\cline{1-12}
\end{tabular}}
\end{center}
\end{table}

\textbf{Ablation: Quantization Precision}. In SPLUT, we uniformly quantize the real-value activations in the aggregation modules during training to control the size of LUTs. 
Since we set $n=4$ for all the LUTs, the prediction accuracy and model scale are mainly determined by $v$. For the spatial lookup blocks, we fix the sampling interval to 16. We change the precision of quantizing the real-value intermediate features, $v_f$, and investigate the influence of it.  
Table~\ref{tab:Sampling Interval} presents the performance of SPLUT-M with different $v_f$. The results indicate that increasing $v_f$ constantly improves the SR accuracy but the model size also increases. When $v_f$ is less than 16, the accuracy improves rapidly. However, SPLUT only has a minor improvement when $v_f$ is greater than 16. Hence we choose $v_f=16$ as the appropriate value which presents appealing SR performance with a relatively small model size. In practice, the quantization precision can be determined by the application scenarios to achieve the flexible model design.   

\begin{table}[t]
\scriptsize
\begin{center}
\caption{Ablation study on LUT number. Extending the depth and width of the SPLUT network both boost the SR performance, which demonstrates the effectiveness and extensibility of the proposed method. }
\label{tab:Layer num}
\renewcommand\arraystretch{1.5} % set height
\setlength{\tabcolsep}{2.5pt}{ % set width
\begin{tabular}{c|c|cc|cc|cc|cc|cc}
\hline
\cline{1-12}
\multirow{2}{*}{Layer   num} & \multirow{2}{*}{Size}      & \multicolumn{2}{c|}{Set5}           & \multicolumn{2}{c|}{Set14}          & \multicolumn{2}{c|}{BSDS100}        & \multicolumn{2}{c|}{Urban100}       & \multicolumn{2}{c}{Manga109}        \\ \cline{3-12} 
                    &          & \multicolumn{1}{c|}{PSNR}  & SSIM   & \multicolumn{1}{c|}{PSNR}  & SSIM   & \multicolumn{1}{c|}{PSNR}  & SSIM   & \multicolumn{1}{c|}{PSNR}  & SSIM   & \multicolumn{1}{c|}{PSNR}  & SSIM   \\ 
                    \hline
                    % \hline
$\rm SPLUT_{1-2}$         & 5M   & \multicolumn{1}{c|}{29.77} & 0.846 & \multicolumn{1}{c|}{27.04} & 0.737 & \multicolumn{1}{c|}{26.56} & 0.696 & \multicolumn{1}{c|}{23.95} & 0.697 & \multicolumn{1}{c|}{26.68} & 0.836 \\
$\rm SPLUT_{1-4}$          & 10M  & \multicolumn{1}{c|}{30.01} & 0.852 & \multicolumn{1}{c|}{27.21} & 0.742 & \multicolumn{1}{c|}{26.66} & 0.700 & \multicolumn{1}{c|}{24.12} & 0.705 & \multicolumn{1}{c|}{27.05} & 0.844 \\
$\rm SPLUT_{1-2-2}$        & 7M   &
\multicolumn{1}{c|}{30.23} & 0.857 & \multicolumn{1}{c|}{27.32} & 0.746 & \multicolumn{1}{c|}{26.74} & 0.704 & \multicolumn{1}{c|}{24.21} & 0.709 & \multicolumn{1}{c|}{27.20} & 0.848 \\
$\rm SPLUT_{1-4-4}$        & 18M  & \multicolumn{1}{c|}{30.52} & 0.863 & \multicolumn{1}{c|}{27.54} & 0.752 & \multicolumn{1}{c|}{26.87} & 0.709 & \multicolumn{1}{c|}{24.46} & 0.719 & \multicolumn{1}{c|}{27.70} & 0.858 \\
\hline
\cline{1-12}
\end{tabular}}
\end{center}
\end{table}

\textbf{Ablation: LUT Number}. We conduct ablation studies on the number of LUTs in each parallel branch to further investigate the extensibility of our SPLUT architecture.
As shown in Table \ref{tab:Layer num}, we compare 4 models with different model depths and widths. 
The model is named according to the number of LUTs in each block. $\rm SPLUT_{1-2}$ represents there are two layers of LUTs. The first layer is the spatial lookup block and the second layer is a query block which contains one $\rm LUT_{WC}$ and one $\rm LUT_{HC}$. For $\rm SPLUT_{1-4}$, the second layer contains two different $\rm LUT_{WC}$ and two different $\rm LUT_{HC}$. In this model, the channel number of intermediate feature maps is $n_{in}=16$. $\rm SPLUT_{1-2-2}$ and $\rm SPLUT_{1-4-4}$ have the similar architectures to $\rm SPLUT_{1-2}$ and $\rm SPLUT_{1-4}$ but have two query blocks in each branch. $\rm SPLUT_{1-2-2}$ and $\rm SPLUT_{1-4-4}$ are actually the same as SPLUT-M and SPLUT-L, respectively. 

In Table~\ref{tab:Layer num}, we observe a huge improvement when the network width increases by comparing the first two rows and the last two rows. This indicates that more LUTs per layer can extract more information by the limited number of channels. In this way, the model can obtain a stronger representation ability and better SR reconstruction performance. 
As the number of query blocks increases, we see $\rm SPLUT_{1-2-2}$ achieves an improvement of 0.46 dB over $\rm SPLUT_{1-2}$ while $\rm SPLUT_{1-4-4}$ outperforms $\rm SPLUT_{1-4}$ by about 0.51 dB on the PSNR performance of Set5. It is inferred that cascading multiple layers of LUTs is very effective in enlarging the receptive fields and boosting recovery abilities. 
Comparing $\rm SPLUT_{1-2-2}$ and $\rm SPLUT_{1-4}$, both models have 5 lookup tables. However, $\rm SPLUT_{1-2-2}$ gains a boost of about 0.22 dB.  This indicates that the network depth is more important than network width in SPLUT. Besides, the comparisons demonstrate the effectiveness of enlarging receptive fields. 

%padding
\begin{table}[t]
\scriptsize
\begin{center}
\caption{Effects of of horizontal and vertical aggregation modules. After removing the aggregation modules, the RF size gets smaller and the restoration ability is degraded. }
\label{tab:padding}
\renewcommand\arraystretch{1.5} % set height
\setlength{\tabcolsep}{2pt}{ % set width
\begin{tabular}{c|cc|cc|cc|cc|cc}
\hline
\cline{1-11}
& \multicolumn{2}{c|}{Set5}           & \multicolumn{2}{c|}{Set14}          & \multicolumn{2}{c|}{BSDS100}        & \multicolumn{2}{c|}{Urban100}       & \multicolumn{2}{c}{Manga109}        \\ \cline{2-11} 
& \multicolumn{1}{c|}{PSNR}  & SSIM   & \multicolumn{1}{c|}{PSNR}  & SSIM   & \multicolumn{1}{c|}{PSNR}  & SSIM   & \multicolumn{1}{c|}{PSNR}  & SSIM   & \multicolumn{1}{c|}{PSNR}  & SSIM   \\ 
\hline
% \hline
SPLUT w/o AM           & \multicolumn{1}{c|}{29.64} & 0.839 & \multicolumn{1}{c|}{26.95} & 0.724 & \multicolumn{1}{c|}{26.08} & 0.676 & \multicolumn{1}{c|}{23.50} & 0.677 & \multicolumn{1}{c|}{25.72} & 0.812 \\
SPLUT       &\multicolumn{1}{c|}{\textbf{30.23}} & \textbf{0.857} & \multicolumn{1}{c|}{\textbf{27.32}} & \textbf{0.746} & \multicolumn{1}{c|}{\textbf{26.74}} & \textbf{0.704} & \multicolumn{1}{c|}{\textbf{24.21}} & \textbf{0.709} & \multicolumn{1}{c|}{\textbf{27.20}} & \textbf{0.848} \\
\hline
\cline{1-11} 
\end{tabular}}
\end{center}
\end{table}

\textbf{Ablation: Aggregation Modules}. Table~\ref{tab:padding} shows the performance comparison between the original SPLUT model and the SPLUT model without aggregation modules, SPLUT w/o AM. From the table we see there is a gap of 0.59 dB between the PSNR performance of SPLUT w/o AM and SPLUT on Set5. The key insight is that we expand the receptive field of the intermediate features by fusing the feature maps padded in opposite directions in the aggregation modules. Thus the features become stronger for the subsequent lookup processes. When aggregation modules are removed, the overall receptive field of the final output is significantly reduced compared with that of the original network, leading to performance degradation correspondingly. The experimental results demonstrate the effectiveness of the proposed horizontal and vertical aggregation modules.

\section{Conclusion}

In this paper, we have proposed a series-parallel lookup table network to achieve efficient image super-resolution. On the one hand, we cascade multiple LUTs to enlarge the receptive field size progressively and enhance the representation capacity of the whole network. On the other hand, we design a parallel architecture to fuse the information of MSB inputs and LSB inputs. By doing so, we compensate for the precision loss caused by quantization when establishing LUTs and improve the prediction accuracy. Comprehensive experiments have demonstrated the effectiveness, efficiency, and flexibility of the proposed method. \par

\subsubsection{Acknowledgement}
This work was supported in part by the National Key Research and Development Program of China under Grant 2017YFA0700802, in part by the National Natural Science Foundation of China under Grant 62125603 and Grant U1813218, in part by a grant from the Beijing Academy of Artificial Intelligence (BAAI).

\bibliographystyle{splncs04}
\bibliography{egbib}

% \begin{document}
% \pagestyle{headings}
% \mainmatter

% \makeatletter
% \renewcommand*{\@fnsymbol}[1]{\ensuremath{\ifcase#1\or *\or \dagger\or \ddagger\or
%   \mathsection\or \mathparagraph\or \|\or **\or \dagger\dagger
%   \or \ddagger\ddagger \else\@ctrerr\fi}}
% \makeatother

\title{Learning Series-Parallel Lookup Tables for Efficient Image Super-Resolution\\Supplementary Material}

\titlerunning{Learning Series-Parallel Lookup Tables for Efficient Image Super-Resolution}

\author{Cheng Ma\inst{1,2,}\thanks{Equal contribution.} \and
Jingyi Zhang\inst{1,2,}\samethanks \and
Jie Zhou\inst{1,2} \and
Jiwen Lu\inst{1,2,}\thanks{Corresponding author.}
}

\authorrunning{C. Ma*, J.Zhang*, et al.}
\institute{Beijing National Research Center for Information Science and Technology, China \and
Department of Automation, Tsinghua University, China \\
\email{macheng17@tsinghua.org.cn; zhangjy20@mails.tsinghua.edu.cn}\\
\email{\{jzhou, lujiwen\}@tsinghua.edu.cn}
}

%******************
\maketitle

\appendix
\section{Details of Mapping Modules}
\begin{figure}[t]
\begin{minipage}{0.68\linewidth}
\centering
\includegraphics[width=\linewidth]{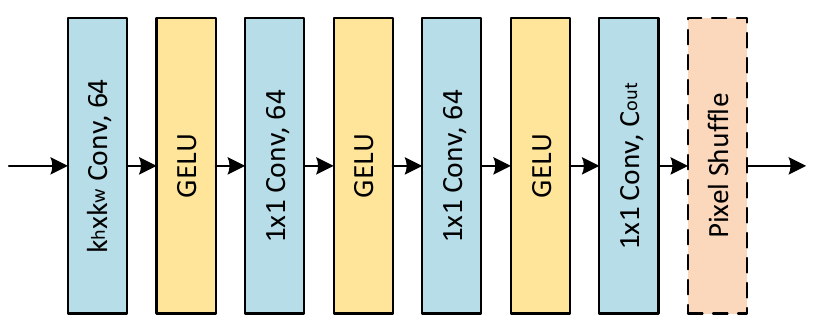}
\caption{
Detailed architecture of the mapping module. Only the last mapping module of each branch contains the pixel shuffle layer. 
}
\label{fig:train}
\end{minipage}
\begin{minipage}{0.3\linewidth}
\centering
\includegraphics[width=\linewidth]{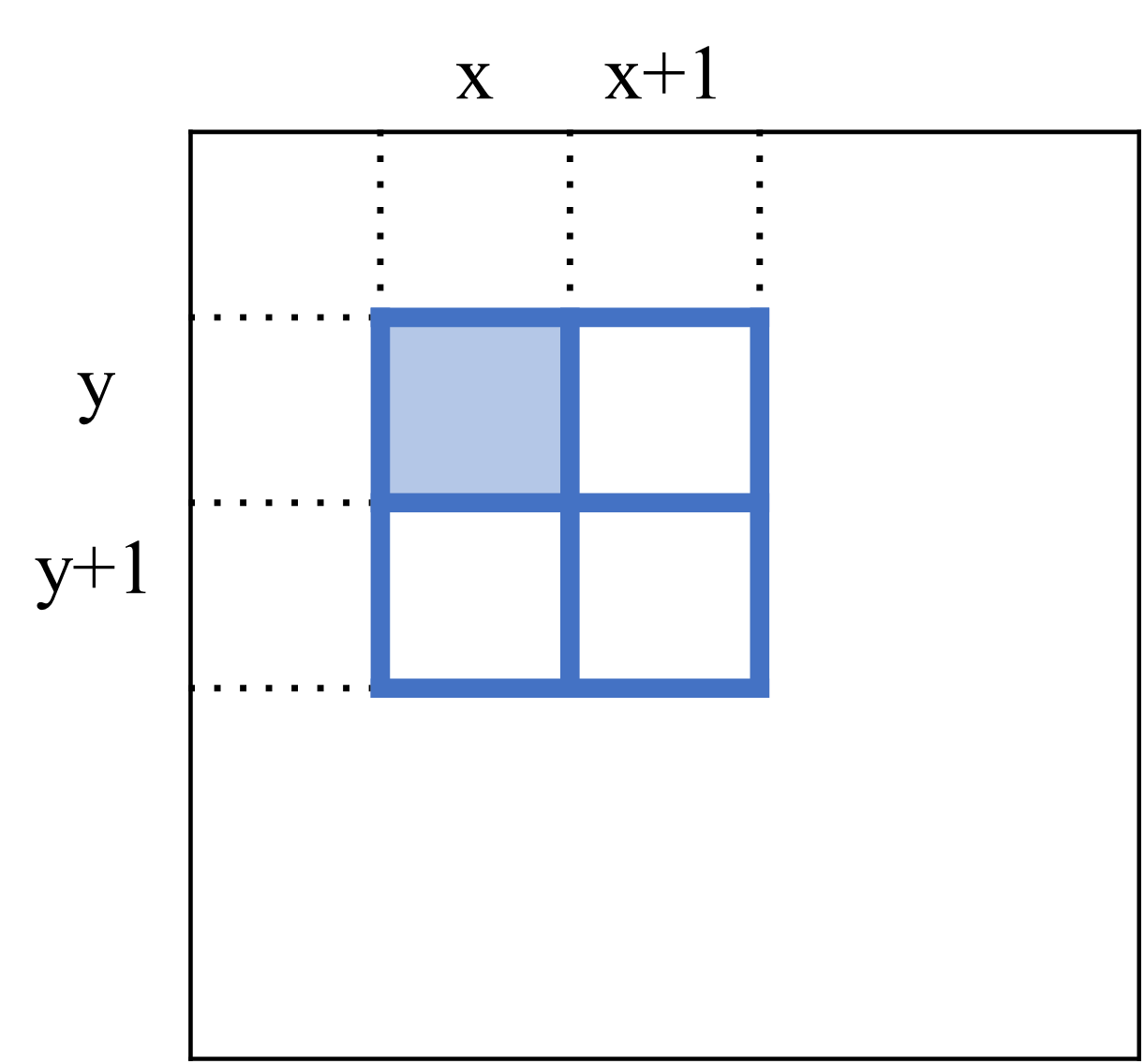}
\caption{ The receptive field of $r=4$ for the pixel colored in blue. 
}
\label{fig:rf4}
\end{minipage}
\end{figure}

In the training phase, we replace the LUTs with mapping modules, which are comprised of a convolutional layer with a kernel size of $k_h\times k_w$, activation layers of GELU~\cite{hendrycks2016gaussian} and $1\times 1$ convolutional layers. The architecture is displayed in Fig.~\ref{fig:train}. 
All convolutional layers output feature maps with 64 channels except the last one. 
The output channel number $C_{out}$ of the last convolutional layer is set to $C_{out}=C_{f}$ for the mapping modules that extract intermediate features and  $C_{out}=C_{SR}=s^2$ for the mapping modules that produce the final SR results. $s$ is the upscaling factor. The consecutive $1\times 1$ convolutions followed by GELU layers strengthen the nonlinearity and representative ability of the mapping modules. In the last query block of each parallel branch, the mapping module contains an additional pixel-shuffle layer~\cite{shi2016real}. It maps the outputs vectors with $s^2$ channels to $s\times s$ output patches to get the final SR images. In the inference phase, we only care about the inputs and outputs of these mapping modules. Therefore, the detailed architecture of the mapping modules does NOT affect the computation complexity of LUTs.

\section{Details of SPLUT-S and SPLUT-L}
\begin{figure*}[t]
\centering
\subfigure[SPLUT-S]{
\centering
\begin{minipage}[b]{0.9\linewidth}
\centering
\includegraphics[width=\linewidth]{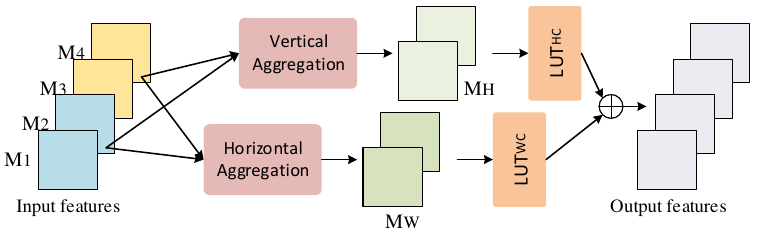}
\end{minipage}
}

\subfigure[SPLUT-L]{
\begin{minipage}[b]{\linewidth}
\centering
\includegraphics[width=\linewidth]{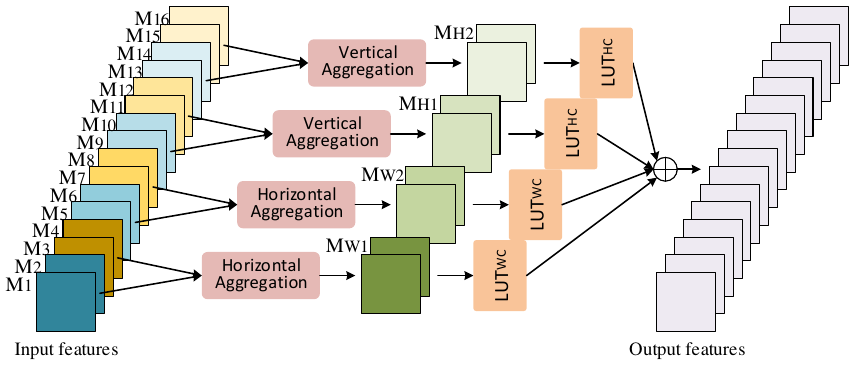}
\end{minipage}
}
\caption{
Details of the proposed query block for (a) SPLUT-S and (b) SPLUT-L. The two models have the same receptive field size. 
}
\label{fig:framework}
\end{figure*}

We design three SPLUT models with different model sizes, namely SPLUT-S, SPLUT-M, and SPLUT-L. The three models have similar architectures. The difference between the three models lies in the channel number of intermediate features $C_f$ and the details of query blocks. We have introduced the details of SPLUT-M in Section 3 of our main paper. SPLUT-S and SPLUT-L have the same overall framework as SPLUT-M but have different query blocks. Note that the three models have the same receptive field size. Here we describe the details of their query blocks, which are shown in  Fig. \ref{fig:framework}. 

%SPLUT-S
SPLUT-S is a small model with $C_f= 4$. In the query block of SPLUT-S, we split 4 intermediate feature maps into 2 groups, each adjacent two in one group. Following SPLUT-M, there is a horizontal aggregation module and a vertical aggregation module in each query block. The two modules both take the above two groups of features as inputs. The obtained features, $M_H$ and $M_W$, are used for the following LUT retrievals. We get the output of the query block by adding the LUT results. 

%SPLUT-L
As for SPLUT-L with $C_f=16$, there are four aggregation modules and four LUTs per query block to increase the capacity of the model. The inputs to the aggregation modules include 8 groups of intermediate features. The first four groups are fed into two vertical aggregation modules and the other four groups are fed into two horizontal modules. The outputs are $M_{H1}$, $M_{H2}$, $M_{W1}$, $M_{W2}$, respectively. We then feed $M_{H1}$ and $M_{H2}$ into two $\rm LUT_{HC}$ and feed $M_{W1}$ and $M_{W2}$ into two $\rm LUT_{WC}$ for further processing.

\section{Details of Two Interpolation Methods}

Here we describe more implementation details of the ablation studies on the parallel network vs. interpolation methods. 
We first introduce the interpolation methods in SRLUT~\cite{jo2021practical} and then provide the details of the two designed interpolation methods. SR approximates the output of an input pattern by interpolating the 4D LUT outputs of the nearest sampled points to the input pattern. 
For a position of $(x,y)$ shown in Fig. \ref{fig:rf4}, the input pixels of $(x,y)$, $(x+1,y)$, $(x,y+1)$ and $(x+1,y+1)$ form an input pattern. SR-LUT implements 4-simplex interpolation for the 4D LUT by exploring the relation of $I_{LSB}^{(x,y)}$, $I_{LSB}^{(x+1,y)}$, $I_{LSB}^{(x,y+1)}$ and $I_{LSB}^{(x+1,y+1)}$. A weighted sum of the retrieval results for the 16 bounding vertices is computed as the final output since SR-LUT has only one layer of LUT. 
Specifically, the nearest sampled points $P_{ijkl}[x][y]$ at the position of $(x,y)$ are calculated as follows:
\begin{eqnarray}
P_{ijkl}[x][y]&=&((I_{MSB}^{(x,y)}+i), (I_{MSB}^{(x+1,y)}+j), (I_{MSB}^{(x,y+1)}+k), (I_{MSB}^{(x+1,y+1)}+l))
\end{eqnarray}
where $i, j, k, l$ are 0 or 1. We can compute the indices for LUT retrieval by these points. SR-LUT selects the contributing bounding vertices by their distances to the input pattern and then use 4-simplex interpolation to compute a weighted sum of their retrieval results according to $I_{LSB}^{(x,y)}$, $I_{LSB}^{(x+1,y)}$, $I_{LSB}^{(x,y+1)}$ and $I_{LSB}^{(x+1,y+1)}$. 

In our methods, we also utilize $I_{LSB}$ for interpolation but we cannot get the final SR output by directly fusing the retrieval results of the first layer of spatial LUT since we still have other following LUTs for retrieval. Since the cascaded LUTs in our model bring a large RF size of $r$, it is intractable to consider all the bounding vertices and simply implement interpolations like SR-LUT due to the computational complexity of $2^r$.
To improve the SR accuracy of this model, we design two interpolation methods for the input images. 
In our first interpolation method, we concatenate the 16 nearest sampled points $P_{ijkl}[x][y]$ of the input patterns of all positions $(x,y)$ to form 16 index maps $P_{0000},P_{0001},...,P_{1111}$. We take $P_{0000},P_{0001},...,P_{1111}$ as the inputs to the cascaded LUTs, which reduce the complexity of $2^r$ to $2^4$. We can get 16 SR results through the whole network. We interpolate the 16 SR results by $I_{LSB}$ using the 4-simplex method and get the final SR output. We call this method tail-layer interpolation. 
In our second method, we feed the 16 index maps to the spatial lookup blocks and get 16 intermediate feature maps. We fuse the 16 feature maps by $I_{LSB}$ and 4-simplex method to get one feature map. By feeding the feature map to the following layers, we get the final SR output. We call this method first-layer interpolation.

%skip
\begin{table}[t]
\scriptsize
\begin{center}
\caption{Ablation study for Skip Connections.}
\label{tab:skip}
\renewcommand\arraystretch{1.5} % set height
\setlength{\tabcolsep}{3pt}{ % set width
\begin{tabular}{c|c|c|c|cc|cc}
\hline
\cline{1-8} 
\multirow{2}{*}{method}&\multirow{2}{*}{SC1} &\multirow{2}{*}{SC2} &\multirow{2}{*}{SC3} & \multicolumn{2}{c|}{Set5}           & \multicolumn{2}{c}{Set14}           \\ \cline{5-8} 
                      &                       &                       & &\multicolumn{1}{c|}{PSNR}  & SSIM   & \multicolumn{1}{c|}{PSNR}  & SSIM   \\ \hline
SPLUT w/o SC3 &\checkmark             & \checkmark            &                       & \multicolumn{1}{c|}{29.95} & 0.8473 & \multicolumn{1}{c|}{27.12} & 0.7366 \\
SPLUT w/o SC2 &\checkmark             &                       & \checkmark            & \multicolumn{1}{c|}{30.16} & 0.8549 & \multicolumn{1}{c|}{27.29} & 0.7452 \\ 
SPLUT w/o SC1 &                      & \checkmark            & \checkmark            & \multicolumn{1}{c|}{30.18} & 0.8562 & \multicolumn{1}{c|}{27.31} & 0.7458 \\
SPLUT&\checkmark             & \checkmark            & \checkmark            & \multicolumn{1}{c|}{30.23} & 0.8567 & \multicolumn{1}{c|}{27.32} & 0.7460 \\
\hline
\cline{1-8} 
\end{tabular}}
\end{center}
\end{table}

\section{Ablation Study for Skip Connections.}

The overall framework of our method is shown in Fig. 2 (a) of the main paper. The first and second skip connections are between the input and output of the spatial LUT block and the first query block. We call the two skip connections ``SC1'' and ``SC2'', respectively. The third skip connection ``SC3'' is between the input image and the output of the last query block. We perform ablation studies to verify the effect of the skip connections. 
As shown in Table \ref{tab:skip}, it can be seen that removing the third skip connection has the greatest impact on SR performance (-0.28dB for Set5). The model with this identity mapping can make the lookup tables pay more attention to the reconstruction of residual information and reduce the difficulty of super-resolution. After removing ``SC3'', the LUTs have to store the information contained in LR inputs, which affects the SR ability of SPLUT. 
The model of SPLUT w/o SC1 has a similar performance to SPLUT w/o SC2. They both have a performance degradation compared to SPLUT. This indicates that the skip connections can help maintain the feature precision and improve inference accuracy. In practice, it is very efficient to implement skip connections with simple addition operations.

\section{Ablation Study for Training Strategy}
To analyze the effectiveness of the jointly training strategy and measure the function of each branch, we conduct experiments on multiple models with different training strategies. 
We remove the MSB branch and train the LSB branch with $I_{LSB}$ as inputs from the scratch to get the SPLUT-LSB model. In this model, $I_{MSB}$ is upsampled by nearest-neighbor interpolation and is added to the network output to form the SR results. We remove the LSB branch and train the MSB branch with $I_{MSB}$ as inputs to get the SPLUT-MSB model. 
We show the corresponding results in Table \ref{tab:A+B}. In the table, ``Interpolation'' represents the results of implementing nearest-neighbor interpolation on $I_{MSB}$.
We see the results of SPLUT-LSB are very close to the nearest-neighbor interpolation. With only $I_{LSB}$ and the interpolated $I_{MSB}$, the network cannot obtain local semantic information and can only generate some meaningless details.
SPLUT-MSB performs much better than SPLUT-LSB and nearest-neighbor interpolation. However, it still has a significant accuracy degradation compared to the original SPLUT model.
The reason may be that the information from $I_{LSB}$ is lost and there are fewer input patterns compared to the full-precision LR images, which aggravates the one-to-many ill-posed nature of SR. 
Furthermore, we design a SPLUT-ST model whose LSB branch and MSB branch are separately trained (ST). 
We first train a model of SPLUT-MSB. Then we fix the parameters of the MSB branch and train the LSB branch. The final results show that the performance of SPLUT-ST is only slightly better than that of SPLUT-MSB, which is similar to the comparison between nearest-neighbor interpolation and SPLUT-LSB. This indicates that the separate training of the LSB branch based on the interpolated $I_{MSB}$ and the pre-trained model of SPLUT-MSB both fail to utilize the information from $I_{LSB}$.
The comparisons demonstrate that only the jointly training strategy can fully exploit the information of the two branches and boost reconstruction accuracy.
%A+B
\begin{table}[t]
\scriptsize
\begin{center}
\caption{Comparison of different training strategies. The results show that jointly training the two parallel branches achieves the best SR performance. }
\label{tab:A+B}
\renewcommand\arraystretch{1.5} % set height
\setlength{\tabcolsep}{2.5pt}{ % set width
\begin{tabular}{c|cc|cc|cc|cc|cc}
 \hline
 \cline{1-11}
\multirow{2}{*}{Layer num}   & \multicolumn{2}{c|}{Set5}           & \multicolumn{2}{c|}{Set14}          & \multicolumn{2}{c|}{BSDS100}        & \multicolumn{2}{c|}{Urban100}       & \multicolumn{2}{c}{Manga109}        \\ \cline{2-11} 
  & \multicolumn{1}{c|}{PSNR}  & SSIM   & \multicolumn{1}{c|}{PSNR}  & SSIM   & \multicolumn{1}{c|}{PSNR}  & SSIM   & \multicolumn{1}{c|}{PSNR}  & SSIM   & \multicolumn{1}{c|}{PSNR}  & SSIM   \\ 
  \hline
Interpolation    & \multicolumn{1}{c|}{26.26} & 0.7380 & \multicolumn{1}{c|}{24.75} & 0.6551 & \multicolumn{1}{c|}{25.04} & 0.6299 & \multicolumn{1}{c|}{22.17} & 0.6160 & \multicolumn{1}{c|}{23.43} & 0.7415 \\
SPLUT-LSB          & \multicolumn{1}{c|}{26.26} & 0.7384 & \multicolumn{1}{c|}{24.80} & 0.6568 & \multicolumn{1}{c|}{25.04} & 0.6305 & \multicolumn{1}{c|}{22.19} & 0.6175 & \multicolumn{1}{c|}{23.49} & 0.7438 \\
SPLUT-MSB         & \multicolumn{1}{c|}{29.54} & 0.8356 & \multicolumn{1}{c|}{26.85} & 0.7217 & \multicolumn{1}{c|}{26.34} & 0.6793 & \multicolumn{1}{c|}{23.90} & 0.6907 & \multicolumn{1}{c|}{26.56} & 0.8316 \\
SPLUT-ST  & \multicolumn{1}{c|}{29.54} & 0.8379 & \multicolumn{1}{c|}{26.92} & 0.7296 & \multicolumn{1}{c|}{26.49} & 0.6912 & \multicolumn{1}{c|}{23.92} & 0.6907 & \multicolumn{1}{c|}{26.64} & 0.8288 \\
SPLUT  &\multicolumn{1}{c|}{\textbf{30.23}} & \textbf{0.8567} & \multicolumn{1}{c|}{\textbf{27.32}} & \textbf{0.7460} & \multicolumn{1}{c|}{\textbf{26.74}} & \textbf{0.7044} & \multicolumn{1}{c|}{\textbf{24.21}} & \textbf{0.7094} & \multicolumn{1}{c|}{\textbf{27.20}} & \textbf{0.8478} \\
\hline
\cline{1-11}
\end{tabular}}
\end{center}
\end{table}

% \bibliography{egbib}
% \end{document}

\end{document}